\pdfoutput=1
\documentclass[aps,prd,reprint,showpacs,showkeys,preprintnumbers]{revtex4-1}

\usepackage[bookmarks=true]{hyperref}
\usepackage{graphicx,epsfig,color,xcolor}
\usepackage[english]{babel}
\usepackage{amssymb,amsfonts,amsmath}
\usepackage{verbatim}

\newcommand{\be}{\begin{equation}} \newcommand{\ee}{\end{equation}}
\newcommand{\bea}{\begin{eqnarray}} \newcommand{\eea}{\end{eqnarray}}

\def\d{\mathrm{d}}

\def\pot{V}
\def\Z{{\cal Z}}
\def\W{{\cal W}}
\def\k{\kappa}

\def\Nf{N_\mt{f}}
\def\Nc{N_\mt{c}}
\def\lym{\lambda_\text{YM}}

\newcommand{\mt}[1]{\textrm{\scriptsize #1}}

\begin{document}

\title{Transport in strongly coupled quark matter}

\author{Carlos Hoyos}
\email{hoyoscarlos@uniovi.es}
\affiliation{Department of Physics and Instituto de Ciencias y Tecnolog\'ias Espaciales de Asturias (ICTEA) \\ Universidad de Oviedo, c/ Federico Garci\'ia Lorca 18, ES-33007 Oviedo, Spain}

\author{Niko Jokela}
\email{niko.jokela@helsinki.fi}
\affiliation{Department of Physics and Helsinki Institute of Physics\\
P.O.~Box 64, FI-00014 University of Helsinki, Finland}

\author{Matti J\"arvinen}
\email{jarvinen@tauex.tau.ac.il}
\affiliation{The Raymond and Beverly Sackler School of Physics and Astronomy \\
 Tel Aviv University, Ramat Aviv 69978, Israel}

\author{Javier G. Subils}
\email{jgsubils@fqa.ub.edu}
\affiliation{Departament de F\'isica Qu\`antica i Astrof\'isica \& Institut de Ci\`encies del Cosmos (ICC), \\ Universitat de Barcelona, Mart\'i Franqu\`es 1, ES-08028, Barcelona, Spain}

\author{Javier Tarr\'io}
\email{javier.tarrio@helsinki.fi}
\affiliation{Department of Physics and Helsinki Institute of Physics\\
P.O.~Box 64, FI-00014 University of Helsinki, Finland}

\author{Aleksi Vuorinen}
\email{aleksi.vuorinen@helsinki.fi}
\affiliation{Department of Physics and Helsinki Institute of Physics\\
P.O.~Box 64, FI-00014 University of Helsinki, Finland}

\begin{abstract}

Motivated by the possible presence of deconfined quark matter in neutron stars and their mergers and the important role of transport phenomena in these systems, we perform the first-ever systematic study of 
different viscosities and conductivities
of dense quark matter using the gauge/gravity duality. Utilizing the V-QCD model, we arrive at results that are in qualitative disagreement with the predictions of perturbation theory, which highlights the differing transport properties of the system at weak and strong coupling and calls for caution in the use of the perturbative results in neutron-star applications.

\end{abstract}

\preprint{HIP-2020-9/TH}
\preprint{ICCUB-20-010}


\maketitle

\section{Introduction}

The first recorded observations of binary neutron star (NS) mergers, including both gravitational wave (GW) \cite{TheLIGOScientific:2017qsa,Abbott:2018exr} and electromagnetic (EM) \cite{GBM:2017lvd} signals, have opened intriguing new avenues for the study of strongly interacting matter at ultrahigh densities. While most of the attention in the field has so far been directed to the macroscopic properties of the stars and correspondingly to the Equation of State (EoS) of NS matter (see e.g.~\cite{Annala:2017llu,Rezzolla:2017aly,Capano:2019eae} and references therein), there exists ample motivation to inspect also transport properties of dense Quantum Chromodynamics (QCD), determined by viscosities and conductivities.
This is in particular due to the fact that 
while near the transition from nuclear to quark matter (QM) the EoSs of the two phases may largely resemble one another, the corresponding transport properties are expected to witness much more dramatic changes,
potentially enabling a direct detection of QM either in quiescent NSs or their binary mergers \cite{Schmitt:2017efp,Most:2018eaw}. In addition, understanding the relative magnitudes of different transport coefficients may turn out useful for the relativistic hydrodynamic simulations of NS mergers \cite{Baiotti:2016qnr,Alford:2017rxf}.

In the QM phase, expected to be found in the cores of massive NSs \cite{Annala:2019puf} and very likely created in NS mergers \cite{Most:2018eaw,Bauswein:2018bma,Chesler:2019osn,Ecker:2019xrw}, very few robust results exist for the QCD contribution to even the most central transport coefficients, including the bulk and shear viscosities and the electrical and heat conductivities. In fact, the only first-principles determination of these quantities in unpaired quark matter dates back to the early 1990s, amounting to a leading-order perturbative QCD (pQCD) calculation  \cite{Heiselberg:1993cr}. Owing to the strongly coupled nature of QM in the density regime relevant for NSs \cite{Kraemmer:2003gd,CasalderreySolana:2011us}, these results are, however, of limited predictive value. A nonperturbative analysis at strong coupling would be required for robust predictions, but the well-known problems that lattice Monte-Carlo simulations face at nonzero baryon densities presently prohibit the use of this standard tool (see e.g.~\cite{deForcrand:2010ys}).

Another complication in the determination of transport quantities in QM has to do with the fact that the physical phase of QCD realized at moderate densities is at present unknown \cite{Alford:2007xm}. While the general expectation is that some type of quark pairing is likely present all the way down to the deconfinement transition at low temperatures \cite{Rajagopal:2000wf}, it is currently unclear which particular phases are realized in Nature. Assuming that the physical moderate-density phase contains at least some nonzero fraction of unpaired quarks, it is, however, often considered a reasonable approximation to first inspect transport coefficients in the somewhat simpler case of unpaired QM \cite{Schmitt:2017efp}.

In the paper at hand, we approach the most important transport properties of dense unpaired QM with the only first-principles machinery currently capable of describing strongly coupled quantum field theories at high baryon density: the gauge/gravity duality or, in short, holography. This correspondence provides a link between classical gravity and strongly coupled quantum field theories, relating their observables in a detailed manner (see e.g.~\cite{CasalderreySolana:2011us,Ramallo:2013bua,Brambilla:2014jmp} for reviews). In the context of heavy-ion physics, many important insights have been gained through the study of questions that are difficult to address with traditional field theory machinery, such as the details of equilibration dynamics (see e.g.~the recent review \cite{Berges:2020fwq}). In addition, the conjectured lower limit of the shear-viscosity-to-entropy ratio, $\eta/s\geq 1/(4\pi)$, has hinted towards universality in strongly coupled systems, which has had a profound effect on many subfields of theoretical physics \cite{Kovtun:2003wp}. 

In the context of NS physics, promising progress has  been achieved in applying holographic methods to 
determining the EoS of QCD matter
\cite{Hoyos:2016zke,Hoyos:2016cob,Ecker:2017fyh,Fadafa:2019euu,Ishii:2019gta}, but no attempts have been made to analyze transport in strongly coupled dense QM. Here, we shall take the first steps in this direction by utilizing one of the most highly developed 
``bottom-up'' frameworks (i.e., with no known string theory origin)
designed to mimic a gravity dual of QCD at finite temperature and density \footnote{Note that at low densities, transport has been studied in the IHQCD and V-QCD frameworks in~\cite{Gursoy:2009kk,Gursoy:2010aa,Iatrakis:2014txa}.}, the Veneziano limit of the Improved Holographic QCD (IHQCD) model, V-QCD  \cite{Jarvinen:2011qe,Alho:2013hsa,Jokela:2018ers}. Interestingly, the results obtained for the most important transport coefficients of dense QM are in qualitative disagreement with known perturbative predictions, calling for caution in the application of the latter results to phenomenological studies of NSs.

For comparison and completeness, we shall contrast our V-QCD results not only to pQCD but also to one of the most widely studied top-down holographic models describing deconfined quark matter, i.e.~the D3-D7 system in its unbackreacted limit \cite{Kruczenski:2003be,Kobayashi:2006sb}, implying we neglect the corrections from flavors. This corresponds to the quenched approximation in field theory language \cite{Nunez:2010sf}.
The latter provides reliable results for quantities that can be derived from the free energy \cite{Karch:2008uy}, which in the present case translates to the shear viscosity, obtainable via the relation $\eta/s=1/(4\pi)$. Other physical quantities are on the other hand expected to receive corrections from backreacted (unquenched) quark matter, which we shall indeed witness in our results.

\section{Setup}

As explained above, we approach the description of dense strongly coupled matter via the string-theory inspired V-QCD model~\cite{Jarvinen:2011qe,Alho:2013hsa,Jokela:2018ers}. The way we have set up our analysis is, however, more general, and can accommodate other holographic models as well, such as the probe-brane limit of the D3-D7 system \cite{Kobayashi:2006sb,Itsios:2016ffv}. Both setups have been thoroughly applied to the study of the bulk properties of NS matter, and we refer the interested reader to Refs.~\cite{Hoyos:2016zke,Annala:2017tqz,Jokela:2018ers,Chesler:2019osn,Ecker:2019xrw,Fadafa:2019euu}.

A significant difference between two holographic models lies in how the effects of the flavor sector are treated. V-QCD has these effects systematically built in, whereas the 
D3-D7 system treats the flavors as a probe
(note, however, that for massless quarks 
refs.~\cite{Bigazzi:2009bk,Bigazzi:2011it,Bigazzi:2013jqa,Faedo:2016cih,Faedo:2017aoe} have gone beyond the probe approximation). Despite this difference, we can define the two models via the same gravitational action consisting of two terms $S_\mt{total} = S_\mt{g} + S_\mt{f}$, where
\begin{eqnarray}
S_\mt{g} &=& N_c^2 M_\text{Pl}^3\int \d^5 x \sqrt{-g} \left( R - \frac{1}{2} \partial_\rho \phi \, \partial^\rho\phi - \pot(\phi) \right) \label{Eq.ActionDefs} \\
S_\mt{f} & =& -N_f N_c M_\text{Pl}^3 \int \d^5 x \Z(\phi,\chi) \sqrt{-\det \left(\Gamma_{\mu\nu} \right) } \label{Eq.ActionWithTachyon}\ .
\end{eqnarray}
Here, $g$ is the determinant of the metric $g_{\mu\nu}$, $M_\text{Pl}$ denotes the rescaled five-dimensional Planck mass, $R$ is the Ricci scalar, and we have defined \be
 \Gamma_{\mu\nu} =  g_{\mu\nu} + \k(\phi,\chi)  \partial_\mu \chi  \partial_\nu\chi + \W(\phi,\chi)  F_{\mu\nu}\ .
\ee
The potentials and couplings $V(\phi)$, $\Z(\phi,\chi)$, $\k(\phi,\chi)$, and $\W(\phi,\chi)$ will be different functions in D3-D7 and in V-QCD as we shall discuss below.

Of the two parts of the action, $S_\mt{g}$ is related to the glue sector of a gauge theory with rank $\Nc$. The scalar field $\phi$ is identified with the dilaton, which according to the holographic dictionary is dual to the Yang--Mills running coupling constant. For V-QCD, following the IHQCD model \cite{Gursoy:2007cb,Gursoy:2007er}, the potential $V$ in~\eqref{Eq.ActionDefs} is chosen to reproduce the known physics of Yang-Mills theory upon comparison with perturbative results and lattice data~\cite{Gursoy:2009jd,Panero:2009tv,Jokela:2018ers}. In the D3-D7 model, $S_\mt{g}$ is on the other hand determined by the closed string sector of type IIB supergravity and $V=-12/L^2$. To fix units, we demand that the asymptotically $AdS$ space has unit radius $L=1$.

At the same time, the physics of $N_f$ flavors of fundamental quarks is captured by the Dirac--Born--Infeld (DBI) action $S_\mt{f}$, where the tachyon field $\chi$ is dual to the chiral condensate $\bar qq$, whose boundary value is related to the masses of the quarks~\footnote{For V-QCD, the flavor setup is based on~\cite{Bigazzi:2005md,Casero:2007ae,Iatrakis:2010zf,Iatrakis:2010jb}}. We set the quark masses to zero in V-QCD, but keep them non-vanishing in the D3-D7 model to achieve the breaking of conformal symmetry. In the D3-D7 model, a quark mass corresponds to the energy necessary to introduce an additional quark over the ground state, implying that it corresponds to a constituent quark mass, with a value of the order of $1/N_c$ times the baryon mass \cite{Hoyos:2016zke}.
The field strength $F_{\mu\nu}=\partial_\mu A_\nu-\partial_\nu A_\mu$ on the other hand provides the dynamics for the U(1)$_\mt{B}$ gauge field $A_\mu$ corresponding to the conserved baryonic charge of the dual field theory. 

\begin{figure*}[t!]
\center
\includegraphics[width=0.45\textwidth]{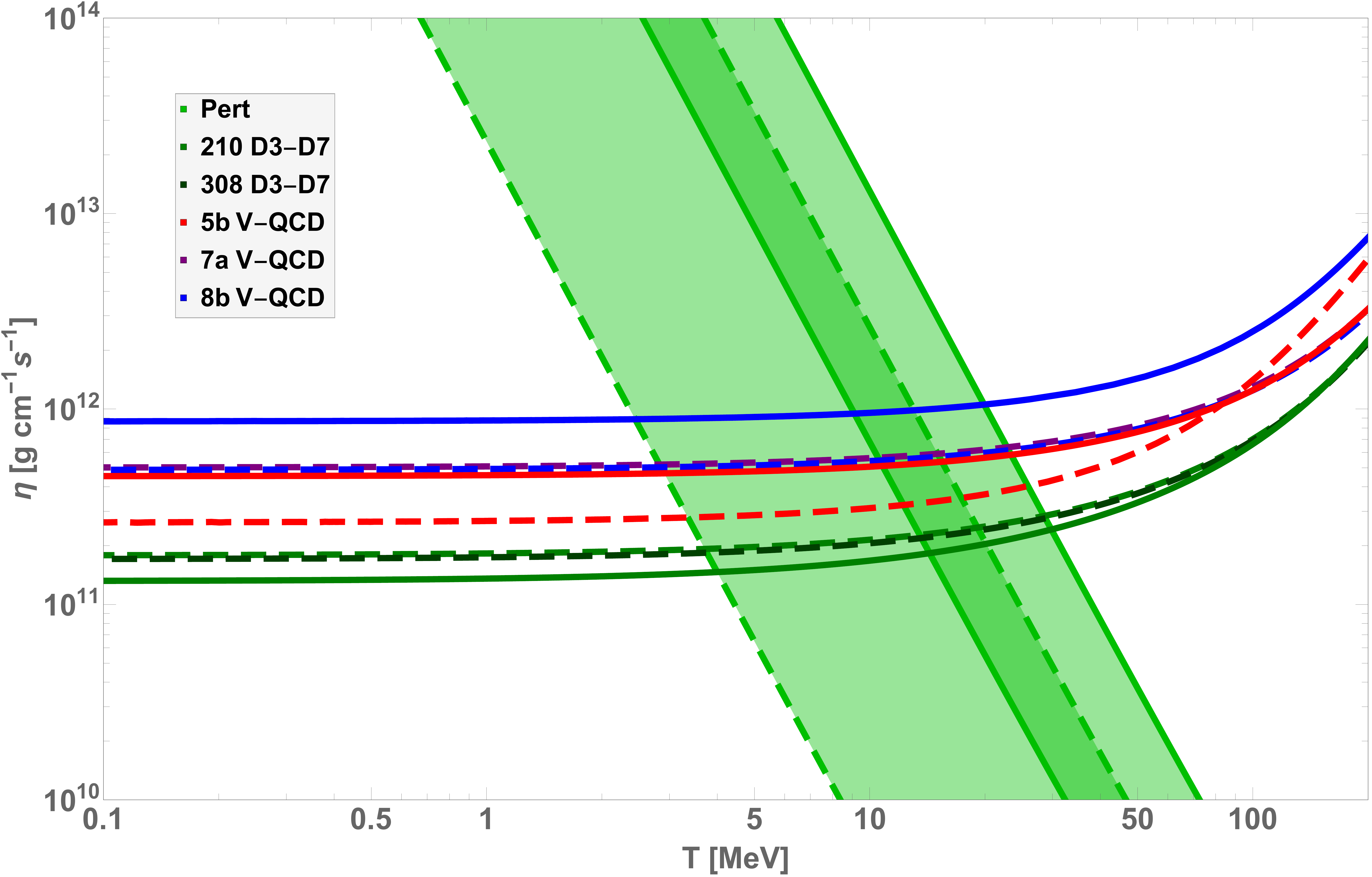}
$\;\;\;\;\;$ 
\includegraphics[width=0.45\textwidth]{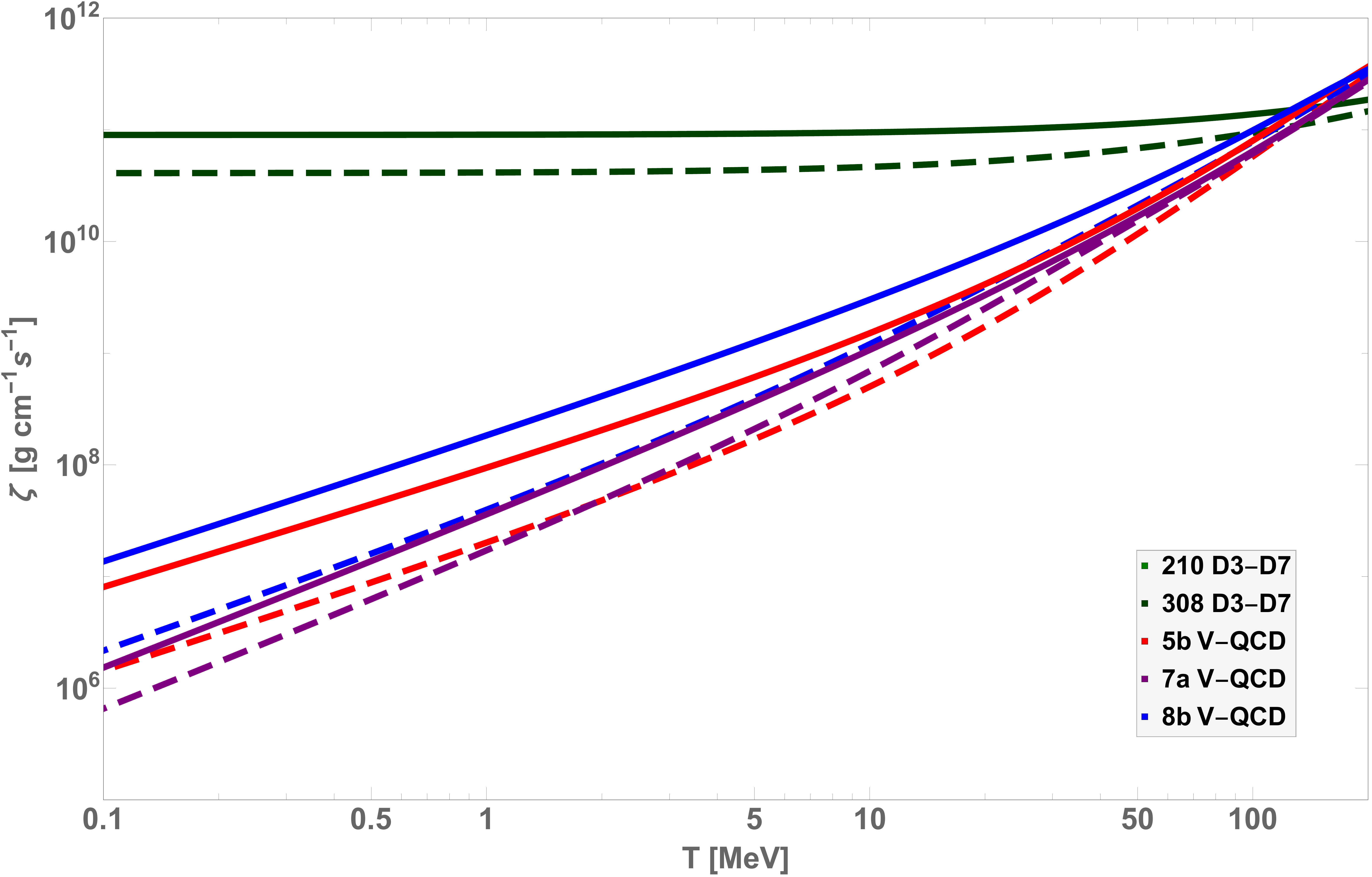}
\caption{Shear (left) and bulk (right) viscosities as functions of temperature for $\mu=450\,\operatorname{MeV}$ (dashed lines) and $\mu=600\,\operatorname{MeV}$ (solid lines). The filled bands on the left correspond to the pQCD results (upper band obtained for $\mu=600$ MeV, lower for 450 MeV), and they have been generated by varying the parameter $x$ inside $\alpha_s$ (see the main text for the definition) inside the interval $1/2\leq x\leq 2$.}
 \label{fig:visc}
\end{figure*}

In the V-QCD model, the functions $\Z$, $\k$, and $\W$ are fixed to reproduce the desired features of QCD (such as confinement and asymptotic freedom) both for weak \cite{Gursoy:2007cb,Jarvinen:2011qe} and strong \cite{Gursoy:2007er,Jarvinen:2011qe,Arean:2013tja,Arean:2016hcs,Jarvinen:2015ofa,Ishii:2019gta} Yang-Mills coupling. Moreover, the potentials were tuned to match with lattice data for the EoS at small chemical potentials \cite{Jokela:2018ers}, see Appendix~\ref{app:vqcd}. In this way, the model in effect extrapolates the lattice results to the regime relevant for NS cores. We also determine $M_\text{Pl}$ by using lattice data and set $N_f/N_c=1$, and furthermore employ the couplings of the fits \textbf{5b}, \textbf{7a}, and \textbf{8b} given in Appendix~A of~\cite{Jokela:2018ers}. 

For the D3-D7 model, supergravity implies the simple relations 
\begin{equation}
M_\text{Pl}^3=\frac{1}{8\pi^2},\, \Z=\frac{\lym}{2\pi^2}\cos^3\chi,\, \W=\frac{2\pi}{\sqrt{\lym}},\,  \k=1,
\end{equation}
where $\lym=g_{_{YM}}^2 N_c$ is the 't Hooft coupling (with $g_{_{YM}}$ the Yang-Mills coupling) of the dual field theory. Following \cite{Hoyos:2016zke}, we fix $\lym\simeq 10.74$ so that the pressure matches the Stefan-Boltzmann value at asymptotically large chemical potentials.

The holographic determination of the thermodynamic quantities and transport coefficients is briefly reviewed in Appendix~\ref{app:setup}. To translate the resulting expressions into numerical results, we need to choose, among other things, the characteristic perturbative energy scale $\Lambda_\text{UV}$ for V-QCD. We use the values $\Lambda_\text{UV}=226.24,\, 210.76,\, \text{and}\, 156.68\, \operatorname{MeV}$ for the fits \textbf{5b}, \textbf{7a}, and \textbf{8b} in the V-QCD model, following the choices made in \cite{Jokela:2018ers,Chesler:2019osn}. The quark mass in the D3-D7 model is on the other hand given  two values --- $M_q=210.76\, \operatorname{MeV}$ for direct comparison with the V-QCD potential  \textbf{7a} \footnote{Recall that these parameters, $M_q$ for D3-D7 and $\Lambda_\text{UV}$ for V-QCD, both control how the conformal symmetry is broken. Their values must therefore be comparable, even though not necessarily exactly the same.}, and $M_q=308.55\, \operatorname{MeV}$ following the logic of ref.~\cite{Hoyos:2016zke} --- which are used to (partially) probe the systematic uncertainties of this model. 

An important point to note is that in both of our models, we consistently work with three mass-degenerate quark flavors assuming beta equilibrium, whereby all quark flavors share the same chemical potential $\mu=\mu_\text{B}/3$. This implies the absence of electrons in the system, which would only change upon taking flavor-dependent masses into account.

\section{Viscosities}

The shear and bulk viscosities of dense QCD matter describe the resistance of the system to deformations. They become relevant in settings where NSs are either strongly deformed or their interiors taken out of thermal equilibrium, both of which occur in different stages of binary NS mergers \cite{Alford:2017rxf,Fujibayashi:2017puw}. In addition, viscosities play a role in determining the damping of unstable r-modes in rapidly rotating stars \cite{Andersson:1997xt,Friedman:1997uh,Andersson:2000mf,Arras:2002dw,Alford:2010gw,Haskell:2015iia}.

Viscosities appear in contributions to stress forces due to an inhomogeneous motion of the fluid. Letting $v_i$, $i=1,2,3$, be the components of the velocity of a fluid moving at low 
velocities, the resulting stress becomes
\begin{equation}
    T_{ij}=-\eta\  \left(\partial_{i}v_{j}+\partial_{j}v_{i}\right) -\left(\zeta-\frac{2}{3}\eta \right)\delta_{ij}\partial_k v^k\ ,
\end{equation}
where $\eta$ and $\zeta$ are the shear and bulk viscosities, respectively.

The values obtained from the holographic models are plotted in Fig.~\ref{fig:visc} (left) together with the pQCD result for unpaired quark matter \cite{Heiselberg:1993cr,Schmitt:2017efp},
\begin{equation}
    \eta\approx 4.4\times 10^{-3} \frac{\mu^2 m_\text{D}^{2/3}}{\alpha_s^2T^{5/3}}\ .
\end{equation}
In evaluating this expression, we have used the one-loop Debye mass $m_\text{D}^2=2\alpha_s(N_f \mu^2+(2N_c+N_f)\pi^2 T^2/3)/\pi$ (see e.g.~\cite{Vuorinen:2003fs}) and the two-loop strong coupling $\alpha_s$, related to the QCD gauge coupling via $\alpha_s=g_{_\text{QCD}}^2/(4\pi)$. Following typical conventions in the literature (see e.g.~\cite{Kurkela:2009gj,Kurkela:2016was} and the discussion in the beginning of Sec.~7 in \cite{Ghiglieri:2020dpq}), the renormalization scale is set to $\bar{\Lambda}=x\sqrt{(2\pi T)^2+(2\mu)^2}$, where $x$ is a parameter that parametrizes the renormalization scale dependence of the pQCD result and is varied between the values 1/2 and 2. Finally, the QCD scale $\Lambda_\text{QCD}$ is given the value 378 MeV, obtained by demanding that $\alpha_s(\bar{\Lambda}=2\, \textrm{GeV})=0.2994$ \cite{Amsler:2008zzb}.

We observe that at high temperatures the shear viscosity of the strongly coupled fluid is qualitatively larger than the perturbative result, while at low temperatures it approaches a constant with the crossing of the holographic and pQCD results taking place around $T\sim 5-50\,\operatorname{MeV}$. We also note that in agreement with our naive expectation, both holographic models give comparable results, as the D3-D7 calculation is not hampered by problems related to backreaction in this case.

\begin{figure*}[t!]
\center
\includegraphics[width=0.45\textwidth]{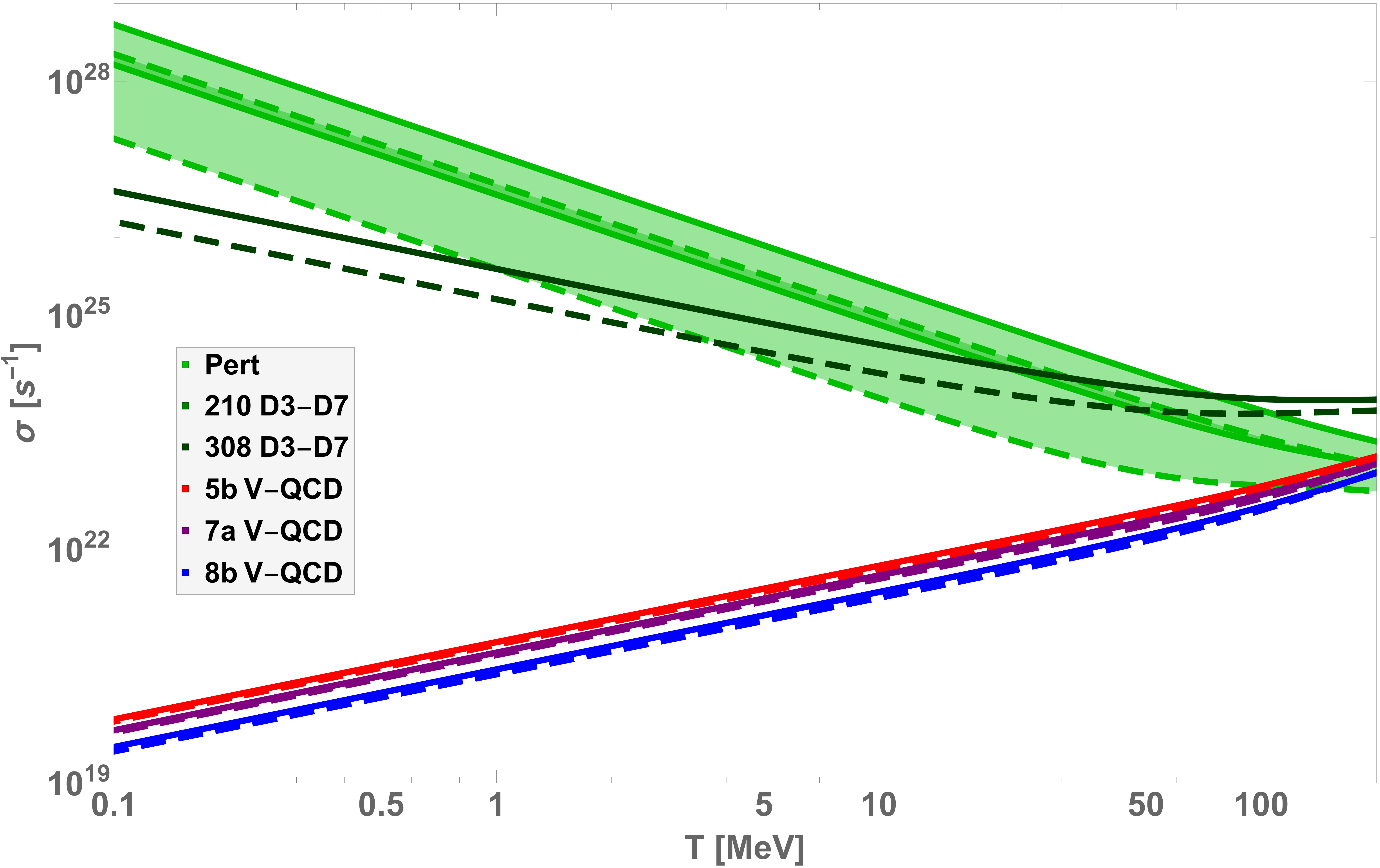}
$\;\;\;\;\;$ 
\includegraphics[width=0.45\textwidth]{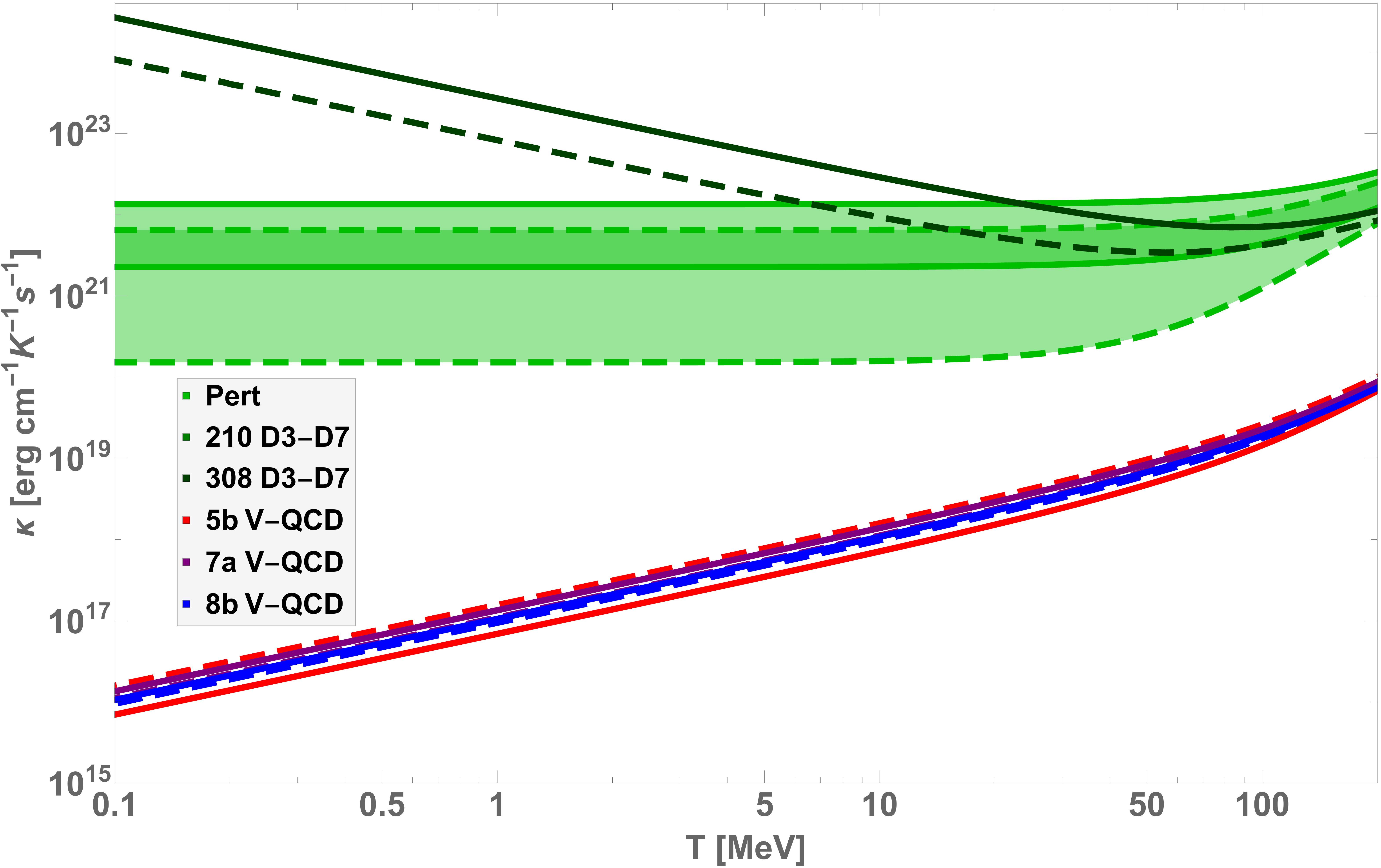}
\caption{Electrical (left) and thermal (right) conductivities as functions of temperature for $\mu=450\,\operatorname{MeV}$ (dashed lines) and $\mu=600\,\operatorname{MeV}$ (solid). The filled bands again correspond to the pQCD results for $1/2\leq x\leq 2$, with the $\mu=600\,\operatorname{MeV}$ bands slightly above the 450 MeV ones.}
 \label{fig:cond}
\end{figure*}

In addition, we have determined the QCD contribution to the bulk viscosity $\zeta$, which  however represents only a subdominant contribution to r-mode damping \footnote{The dominant one originates from chemical equilibration through weak processes, $\zeta \sim 10^{24}-10^{29}\,\operatorname{g\, cm^{-1}s^{-1}}$ \cite{Alford:2010gw,Schmitt:2017efp}. A quantitative analysis of this effect is unfortunately outside the scope of our work.}. The corresponding results are shown in Fig.~\ref{fig:visc} (right), where we observe that in V-QCD $\zeta$ is highly suppressed in comparison with $\eta$ at low temperatures, but reaches a value of around $10\%$ of the shear viscosity at high temperatures. The D3-D7 model on the other hand leads to a very flat curve in the range of temperatures studied, but the validity of this model is clearly questionable due to the flavor quenching. In leading-order pQCD, the bulk viscosity finally vanishes when the quark masses are negligible compared to the chemical potentials, and remains small even at high temperatures \cite{Arnold:2006fz} (note however that nonperturbative effects are expected to increase the value somewhat \cite{Kharzeev:2007wb,Wang:2011ur}), which explains the lack of a pQCD curve in this figure.

\section{Conductivities}

A newly formed NS undergoes a cooling process mainly through the emission of neutrinos from its interior. The neutrinos transport heat to the surface, where the energy is emitted as radiation. A closer inspection of this process shows that the thermal evolution of the star depends on several quantities, including the heat conductivity that determines the heat flux to the surface and the electrical conductivity that determines the magnitude of Joule heating through the decay of magnetic fields \cite{Yakovlev:2004iq,Page:2005fq,Vigano:2013lea}. In the postmerger phase of an NS binary merger, the electrical and thermal conductivities are furthermore relevant for equilibration and the evolution of magnetic fields \cite{Alford:2017rxf,Harutyunyan:2018mpe}. Finally, these quantities may in principle prove useful in distinguishing between different phases of QCD through the observation of thermal radiation.

In the strongly coupled theories that we study in the present work, matter resides in a state that can be described as a relativistic fluid. This implies that the electrical and thermal conductivities are not independent, but are determined by a single coefficient $\sigma$, defined by the constitutive relation of the current arising from a gradient of the chemical potential,
\begin{equation}
    J_x=-\sigma\partial_x\left(\frac{\mu}{T}\right)\ .
\end{equation}
The electrical conductivity, defined as the ratio between the current and the electric field $E_x$, can be shown to take the form, see Appendix~\ref{app:cond},
\begin{equation}
    \sigma^{xx}=\frac{J^x}{E_x}=\frac{\varepsilon+p}{Ts}\sigma\ ,
\end{equation}
where $\varepsilon$, $p$, and $s$ are the energy density, pressure, and entropy density, respectively. Moreover, the thermal conductivity, defined as the ratio between the heat current $Q^x$ and the temperature gradient, becomes
\begin{equation}\label{eq:kappaxx}
    \kappa^{xx}=\frac{Q^x}{-\partial_x T}=\frac{\mu}{T}\frac{\varepsilon+p}{\rho}\sigma =\frac{\mu\, s}{\rho} \,\sigma^{xx}\ ,
\end{equation}
where $\rho$ is the charge density. It should be stressed that these expressions hold only for a steady state, where the gradients of the temperature and chemical potential balance each other, so that the transport does not occur via convection.

With the above results established, we see that it suffices to compute the electrical conductivity in the holographic models. Our results for the two conductivities are displayed in Fig.~\ref{fig:cond} together with the perturbative results for unpaired quark matter \cite{Heiselberg:1993cr,Schmitt:2017efp} 
\begin{equation}
    \sigma^{xx}\approx 0.01\frac{\mu^2 m_\text{D}^{2/3}}{\alpha_s T^{5/3}}\ , \ \kappa^{xx}\approx 0.5 \frac{m_\text{D}^2}{\alpha_s^2} \ .
\end{equation}
Here, we continue to use the same values for $m_\text{D}$ and $\alpha_s$ as listed in the previous section, and the electrical conductivity is given in units of $e^2/(\hbar c)$.

We observe that both in the perturbative and D3-D7 calculations the conductivities either decrease with temperature or are largely independent of it, while in the V-QCD model they are increasing functions of $T$. This qualitative discrepancy can be easily understood
as a suppression of momentum dissipation of charged particles coming from the assumptions of these models: in the pQCD case, the coupling is assumed  small by construction, whereas in the D3-D7 case, only the leading order effect in an $\Nf/\Nc$  expansion is retained \cite{Karch:2007pd}.
There are strong reasons to expect that properly including backreaction in this holographic model would bring the D3-D7 result into at least qualitative agreement with the V-QCD one (see e.g.~the discussion in \cite{Tarrio:2013tta}).

\section{Discussion}

A first-principles microscopic determination of the fundamental properties of dense QCD matter is a notoriously difficult problem, not least due to the strongly coupled nature of the system at phenomenologically relevant energies. In recent years, the gauge/gravity duality has shown considerable promise as a potential tool, as it allows approaching the problem from an angle complementary to traditional field theory methods; indeed, promising results have been obtained for many bulk thermodynamic quantities, leading to predictions for observables such as the NS mass-radius relation and the phase diagram of the theory \cite{Hoyos:2016zke,Jokela:2018ers,Chesler:2019osn}. 

In the paper at hand, we have used the holographic machinery to tackle a more challenging class of physical quantities characterizing the response of the medium to external perturbations. In particular, we have studied the behavior of the transport coefficients most relevant for the physics of NSs and their mergers, i.e.~the shear and bulk viscosities and the thermal and electrical conductivities. All  these quantities have been evaluated in a highly developed bottom-up framework mimicking a gravity dual for QCD, V-QCD, and the corresponding results subsequently compared to those from the D3-D7 probe brane setup and perturbative QCD \cite{Heiselberg:1993cr}.

Our main results are depicted in Figs.~\ref{fig:visc} and \ref{fig:cond}. Inspecting these plots, an issue that stands out immediately is the stark contrast between the V-QCD and pQCD predictions for all quantities, for which both results are available: in V-QCD, transport coefficients typically increase with  temperature, while a qualitatively different behavior is predicted by pQCD. As a result, in the $T=0$ limit V-QCD coefficients are comparatively strongly suppressed. In addition, while we observe qualitative agreement between the V-QCD and D3-D7 predictions for the shear viscosity, the same is not true for the other three quantities studied. These observations clearly call for physical interpretation.

As flavors are quenched in the D3-D7 model, the main dissipative effect affecting quark matter is due to drag by the thermal plasma \cite{Gubser:2006bz,Herzog:2006gh}. The increasing trend of the electrical conductivity at lower temperatures reflects the decreasing drag force, which is however expected to be cut off by radiation effects not captured by the D3-D7 analysis 
\cite{Mikhailov:2003er,Chernicoff:2008sa,Karch:2007pd}. Furthermore, at very low temperatures, the quenched approximation is expected to break down altogether \cite{Bigazzi:2011it,Bigazzi:2013jqa}. The V-QCD model on the other hand does not suffer from these issues, as flavors are unquenched, and consequently the trends exhibited by  both conductivities can be expected to reflect the true behavior of these quantities  in strongly coupled unpaired quark matter.

For the viscosities, the situation is somewhat different. In both holographic setups, the shear viscosity is proportional to a quantity derived from the free energy, which leads to a fair agreement between the two predictions. On the other hand, in the pQCD calculation flavor contributions give rise to a strong increase of the quantity in the $T\to 0$ limit, leading to a stark disagreement with the holographic results. In contrast, for the bulk viscosity, for which no pQCD result is available, we witness a marked sensitivity of the result to the pattern in which conformal invariance is broken (by a running coupling vs.~quark masses) in our holographic models, which essentially invalidates the prediction of the D3-D7 model for this quantity.

In summary, we have seen dramatic differences arise between the predictions of a strongly coupled unquenched theory and its perturbative or quenched approximations for the transport properties of dense quark matter. This observation clearly calls for significant caution in the application of the perturbative results in any phenomenological study within NS physics, and highlights the necessity of further developing the holographic approach to the problem.

\vspace{0.3cm}

\begin{acknowledgments}
{\em  Acknowledgments}$\;$ We thank Christian Ecker, Umut G\"ursoy, Jacob Sonnenschein, and Andreas Schmitt for useful discussions and comments on the manuscript.  The work of C.H.~has been partially supported by the Spanish grant PGC2018-096894-B-100 and by the Principado de Asturias through the grant GRUPIN-IDI/2018 /000174; the work of N.J., J.T., and A.V.~by the European Research Council, grant no.~725369, and by the Academy of Finland grants no.~1322307 and 1322507; and the work of M.J.~in part by a center of excellence supported by the Israel Science Foundation, grant no.~2289/18. J.G.S.~acknowledges support from the FPU program, Fellowships FPU15/02551 and EST18/00331, and has been partially supported by grants no.~FPA2016-76005-C2-1-P, FPA2016-76005-C2-2-P, SGR-2017-754, and MDM-2014-0369. The authors finally acknowledge support from CNRS through the PICS program as well as from the Jenny and Antti Wihuri Foundation.
\end{acknowledgments}

\appendix


\section{Technical details of the holographic calculation}\label{app:setup}

For the metric, the Ansatz corresponding to a homogeneous, rotationally-invariant background reads
\be\label{Eq.Ansatz}
\d s^2  =  g_{tt}(r) \d t^2 + g_{xx}(r) \d \vec x ^2 + g_{rr}(r) \d r^2\ ,
\ee
where $r$ is the holographic radial coordinate. In addition, we use the fact that the gauge potential and scalar fields are only functions of the holographic radial coordinate, $A_t=A_t(r),\;\phi=\phi(r),\;\chi=\chi(r)$. Owing to the fact that we are interested in the deconfined phase of the field theory, there is a black hole of radius $r_\text{H}$ in the interior, with $r_\text{H}$ determined by the condition $g_{tt}(r_\text{H})=0$. This is utilized e.g.~when we write down closed formulas for the transport coefficients in terms of the potentials and fields evaluated at $r=r_\text{H}$, with subscript $H$ referring to this evaluation.

We also recall generic expressions for the temperature $T$, entropy density $s$, and charge density $\rho$:
\bea
    4\pi T & = & \left|\frac{d}{dr}\sqrt{-\frac{g_{tt}(r)}{g_{rr}(r)}}\right|_{r=r_\text{H}},\ \ s=4\pi N_c^2 M_\text{Pl}^3(g_{xx}^\mt{H})^{3/2} \nonumber\\  
    \frac{4\pi\rho}{s} & = & -\frac{N_f}{N_c}\frac{ \Z_\mt{H}\W_\mt{H}^2 F_{rt}^\mt{H}}{\sqrt{1-\W_\mt{H}^2 (F_{rt}^\mt{H})^2}}\ .
\eea
In contrast, the quark chemical potential $\mu\equiv \mu_\text{B}/N_f$ is determined by the boundary value of the gauge potential, $\mu=A_t|_{\textrm{bdry}}$, 
given the regularity condition $A_t(r_\text{H})=0$. In the D3-D7 system, in the canonical gauge for the radial coordinate $r$, the quark mass is determined from the asymptotic expansion of the scalar \cite{Mateos:2006nu}:
$M_q/T=\lym^{1/2}/2\,\lim_{r\to\infty}r\chi(r)/r_\text{H}$.

The thermodynamic energy density $\varepsilon$ and the pressure $p$ can finally be derived from the thermodynamic relations $\varepsilon+p=Ts+\mu \rho$ and $\partial_T p=s$. Alternatively, $p$ may also be computed by evaluating the action of the holographic model in the on-shell limit. To solve the metric and thermodynamics in the V-QCD setup, we used the Mathematica package available at~\cite{VQCDThermo}.

In many holographic models, the shear viscosity saturates the  Kovtun--Son--Starinets (KSS) bound $\eta/s=1/(4\pi)$ \cite{Kovtun:2003wp}. This is in particular the case for both of our holographic models V-QCD and D3-D7, which allows us to evaluate the quantity with ease. Similarly, to compute the bulk viscosity we use the Eling-Oz formula \cite{Eling:2011ms}
\begin{equation}
    \begin{split}
\frac{\zeta}{\eta}= \left(s\frac{\partial \phi_H}{\partial s}+\rho \frac{\partial \phi_H}{\partial\rho}\right)^2+\frac{N_f}{N_c}c_H\left(s\frac{\partial \chi_H}{\partial s}+\rho \frac{\partial \chi_H}{\partial\rho}\right)^2\ ,  
\end{split}
\end{equation}
where $c_H=\k_\mt{H}\Z_\mt{H}/\sqrt{1-\W_\mt{H}^2(F_{rt}^\mt{H})^2}$.
The conductivities can finally be computed by extending the methods of \cite{Donos:2014cya,Gouteraux:2018wfe} to a generic DBI action. 
The result obtained in this fashion reads \footnote{One can show that Eq.~(\ref{Eq.DCconductivity}) agrees with the original calculation for the D3-D7 model \cite{Karch:2007pd}.}
\begin{equation}
    \label{Eq.DCconductivity}
    \sigma^{xx} = N_f N_c M_\text{Pl}^3\frac{s\,T}{\varepsilon+p} \frac{(g_{xx}^\mt{H})^{1/2} \Z_\mt{H}\W_\mt{H}^2}{\sqrt{1-\W_\mt{H}^2(F_{rt}^\mt{H})^2}} \ .
\end{equation}
It should be noted that this expression corresponds to the conductivity computed under the steady state condition \eqref{eq:notdep}.

\section{Definitions for the V-QCD model}\label{app:vqcd}

In order to define the V-QCD model precisely, we need to specify the numerical values of the model parameters and the various functions of $\phi$ and $\chi$ in the gravitational action.
In the glue sector, the function $V(\phi)$ in Eq. (1) of the letter is conveniently expressed by using the field $\lambda = e^{\sqrt{3/8}\,\phi}$ which is identified as the 't Hooft coupling:
\bea
    V(\lambda) & = & -12\Bigg[1 + V_1 \lambda + V_2 \frac{\lambda^{2}}{1+\lambda/\lambda_0} \nonumber \\
    & & +V_\mathrm{IR} e^{-\lambda_0/\lambda}(\lambda/\lambda_0)^{4/3}\sqrt{\log(1+\lambda/\lambda_0)}\Bigg] \ .
\eea
Here the asymptotics at strong coupling have been chosen such that the model is confining and has  asymptotically linear trajectories for the squared masses of radially excited glueballs~\cite{Gursoy:2007cb,Gursoy:2007er}.
The parameters $V_1$ and $V_2$ are fixed by requiring agreement with the running of the  Yang-Mills coupling (see discussion below), which leads to 
\begin{equation}
    V_1 = \frac{11}{27\pi^2} \ , \qquad V_2 = \frac{4619}{46656 \pi ^4} \ ,
\end{equation}
so that in particular asymptotic freedom is implemented.
The strong coupling parameters $\lambda_0$ and $V_\text{IR}$ are determined by comparing to lattice data~\cite{Panero:2009tv} for the thermodynamics of pure Yang-Mills at large $N_c$~\cite{Gursoy:2009jd,Alho:2015zua,Jokela:2018ers,Alho:2020gwl}:
\begin{equation}
  \lambda_0 = 8\pi^2/3 \ , \qquad V_\mathrm{IR} = 2.05 \ .
\end{equation}

We then discuss the potentials of the flavor sector in Eq.~(2) of the main text. We first write $\Z(\lambda,\chi)= V_{f0}(\lambda)e^{-\chi^2}$ and employ the Ansatz
\begin{align}
     V_{f0}(\lambda) &= W_0 + W_1 \lambda +\frac{W_2 \lambda^2}{1+\lambda/\lambda_0} + W_\mathrm{IR} e^{-\lambda_0/\lambda}(\lambda/\lambda_0)^{2}  & \\
\frac{1}{\W(\lambda)} &=  w_0\left[1 + \frac{w_1 \lambda/\lambda_0}{1+\lambda/\lambda_0} + 
\bar w_0 
e^{-\hat\lambda_0/\lambda}\frac{(\lambda/\hat\lambda_0)^{4/3}}{\log(1+\lambda/\hat\lambda_0)}\right] \  .
\end{align}
Notice that we do not need the function $\kappa(\lambda)$, because it is the kinetic coupling of the tachyon field $\chi$, and for V-QCD we only consider chirally symmetric configurations at zero quark mass so that $\chi=0$. The strong coupling asymptotics of $V_{f0}$ and $\W$ is chosen such that the asymptotics of the spectra of radially excited mesons are linear~\cite{Jarvinen:2011qe,Arean:2013tja}, solutions are regular at zero~\cite{Jarvinen:2011qe,Arean:2013tja} and at finite $\theta$-angle~\cite{Arean:2016hcs} (in particular the geometry ends in a ``good'' kind of singularity in the classification of~\cite{Gubser:2000nd}), 
and the phase diagram at finite chemical potential includes a confining phase~\cite{Ishii:2019gta}. The parameters $W_1$ and $W_2$ are determined by requiring agreement with the perturbative running of the 't Hooft coupling at two loops for QCD in the Veneziano limit of large $N_c$ and $N_f$ with $N_f/N_c=1$, which gives
\begin{equation}
    W_1 = \frac{8+3\, W_0}{9 \pi ^2} \ , \qquad W_2 = \frac{6488+999\, W_0}{15552 \pi ^4} \ .
\end{equation}
The remaining parameters are then fitted to QCD thermodynamics from lattice with 2+1 dynamical quarks as follows. The function $V_{f0}$ controls the flavor contributions to the thermodynamics at zero chemical potential. Therefore we fit~\cite{Jokela:2018ers} the parameters $W_0$ and $W_\mathrm{IR}$ as well as the Planck mass $M_\text{Pl}$ to the interaction measure $(\epsilon-3p)/T^4$~\cite{Borsanyi:2013bia}. Being a coupling of the field strength tensor, the function $\W$ controls the thermodynamics at finite $\mu$. Therefore we fit the parameters $\hat \lambda_0$, $w_0$, $w_1$, and $\bar w_0$ to the lattice QCD data for the baryon number susceptibility $\chi_2 = d^2p/d\mu^2|_{\mu=0}$ with 2+1 dynamical flavors~\cite{Borsanyi:2011sw}. The results for the three fits  \textbf{5b}, \textbf{7a}, and \textbf{8b} used in this letter can be found in Table~2 of~\cite{Jokela:2020piw}.

Finally, the energy scale of the model is a parameter which, as in real QCD, does not appear in the action but is a property of the solutions. It is convenient to specify the scale by studying the asymptotics of the solution in the weak coupling regime. For a different approach for comparison to the QCD scale and running coupling in a holographic framework see~\cite{Brodsky:2010ur}.

We choose a gauge where $g_{tt}(r)g_{rr}(r)=g_{xx}(r)^2$ and 
choose the coordinate value where the coupling vanishes to be $r=0$. The Einstein equations then imply that
\bea
    g_{xx}(r) & = & \frac{\ell^2}{r^2}\left[1+\frac{8}{9 \log r \Lambda_\text{UV}} +\mathcal{O}\left(\frac{1}{(\log r\Lambda_\text{UV})^2}\right)\right]  \\
    \lambda(r) & = & e^{\sqrt{3/8}\, \phi(r)} \nonumber \\
    & = & -\frac{8\pi^2}{3 \log r \Lambda_\text{UV}} -\frac{28 \pi ^2}{27}  \frac{\log(-\log r\Lambda_\text{UV} )}{(\log r \Lambda_\text{UV})^2} \nonumber \\
    & & + \mathcal{O}\left(\frac{1}{(\log r\Lambda_\text{UV})^3}\right) \ ,
\eea
where $\ell = 1/\sqrt{1-W_0/12}$ is the asymptotic AdS radius. These expansions define the energy scale $\Lambda_\text{UV}$. Moreover, the renormalization scale in field theory is identified as 
$\sqrt{g_{xx}(r)}$~\cite{Gursoy:2007cb}, which was used to map these expansions to the perturbative renormalization group flow of QCD in order to determine the coefficients $V_{1}$, $V_2$, $W_1$, and $W_2$ above. Comparing our results to the lattice data for the interaction measure, we find that $\Lambda_\text{UV}=226.24,\, 210.76,\, \text{and}\, 156.68\, \operatorname{MeV}$ for the fits \textbf{5b}, \textbf{7a}, and \textbf{8b}, respectively.

\section{Conductivities in a relativistic fluid}\label{app:cond}

In the  background metric $G_{\mu\nu}$ and in the presence of a background gauge field $A_\mu$, the constitutive relations for the energy-momentum tensor and charge current of a relativistic fluid read
\be
T^{\mu\nu}=(\varepsilon+p) u^\mu u^\nu+p\, G^{\mu\nu}+\tau^{\mu\nu}\ ,\ J^\mu=\rho u^\mu+\nu^\mu\, ,
\ee
where $u^\mu$ is the fluid velocity ($u^\mu u_\mu=-1$), $p$ the pressure, $\varepsilon$ the energy density, and $\rho$ the charge density. The thermodynamic potentials depend on the temperature $T$ and chemical potential $\mu$ that will be the dynamical variables together with the velocity. 

In the above energy-momentum tensor, the terms $\tau^{\mu\nu}$ and $\nu^\mu$ contain derivatives of the fields. We work in the Landau frame, where
\be
u_ \mu \tau^{\mu\nu}=0\ , \ u_\mu\nu^\mu=0\ .
\ee
We note that in the absence of parity violation, the most general derivative terms in the current compatible with the second law of thermodynamics read, to first order in derivatives,
\be
\nu^\mu=\sigma \left( E^\mu-T P^{\mu\nu}\nabla_\nu \left( \frac{\mu}{T}\right)\right)\ .
\ee
Here, $P^{\mu\nu}$ is the projector transverse to the velocity, and  $E^\mu=F^{\mu\nu}u^\nu$ is the electric field. The derivative terms of the energy-momentum tensor will not be relevant in the following.

The dynamics of the fluid is determined by the conservation equations
\be
\nabla_\mu T^{\mu\nu}=F^{\nu\lambda} J_\lambda\ , \ \nabla_\mu J^\mu=0 \ .
\ee
In the absence of sources $G_{\mu\nu}=\eta_{\mu\nu}$, $A_\mu=0$, the energy and charge densities and the pressure are constant, and the fluid is at rest.

We now turn on small homogeneous time-dependent perturbations $h_{\mu\nu}$, $a_\mu$:
\be
G_{\mu\nu}=\eta_{\mu\nu}+ h_{\mu\nu}(t),\ \ A_\mu= a_\mu(t)\, ,
\ee
where $\eta_{\mu\nu}$ is the Minkowski metric. These perturbations will induce a small change in the hydrodynamic variables $T$, $\mu$, and $u^i$ that can be found by solving the hydrodynamic equations to linear order in the sources.
The value of the energy-momentum tensor and the current in the presence of the external sources is then obtained by inserting the solutions for the hydrodynamic variables back in the constitutive relations and expanding to linear order. For the calculation of the conductivities, we can set $h_{00}=h_{ij}=a_0=0$.

The explicit dependence of the currents on the sources turns out to read
\be\label{eq:hydrosol}
\begin{split}
& J_i  = -\frac{\rho^2}{\varepsilon+p} a_i-\rho h_{0i}-\sigma \partial_t a_i \\
& T^0_{\ i}  =-\rho a_i-(\varepsilon+p) h_{0i}\, ,
\end{split}
\ee
while a constant electric field and temperature gradient $\zeta_i=-\partial_i T/T$ correspond to sources linear in time, i.e.
\be
a_i=-t(E_i-\mu \zeta_i)\ , \ h_{0i}=-t \zeta_i \ .
\ee
Introducing these expressions in Eq.~\eqref{eq:hydrosol}, we readily obtain
\be
\begin{split}
&J_i= \frac{\rho}{\varepsilon+p}t\left((\varepsilon+p-\mu \rho)\zeta_i +\rho E_i\right) +\sigma (E_i-\mu\zeta_i) \\
&T^0_{\ i}=t\left[\rho E_i+(\varepsilon+p-\mu\rho)\zeta_i\right] \ .
\end{split}
\ee

One can impose the condition that the fluid remains at rest, $T^0_i=0$ (no convection), by imposing the following relation between the electric field and the gradient of temperature
\be\label{eq:notdep}
\rho\, E_i+T s\, \zeta_i=0\, ,
\ee
where we have used the thermodynamic relation $\varepsilon+p-\mu\rho=T s$. Physically, the forces induced by the electric and temperature gradients compensate each other, so all transport will occur through diffusion.

The charge and heat currents finally become
\be
\begin{split}
&J_i= \sigma E_i-\mu\sigma \zeta_i \\
&Q_i=T^0_{\ i}-\mu J^i=-\mu\sigma  E_i+\mu^2\sigma \zeta_i\, .
\end{split}
\ee
Using Eq.~\eqref{eq:notdep} to solve for $\zeta_i$ in terms of $E_i$, or vice versa, we then obtain for the currents
\be
\begin{split}
&J_i= \sigma\frac{\varepsilon+p}{Ts} E_i \\ 
&Q_i=T^0_{\ i}-\mu J^i=\mu\sigma\frac{\varepsilon+p}{\rho}\zeta_i\ ,
\end{split}
\ee
so that the electrical and  thermal conductivities read
\be
 \sigma^{ij}= \sigma\frac{\varepsilon+p}{Ts}  \delta^{ij} \ , \ \kappa^{ij}=\frac{\mu}{T}\sigma\frac{\varepsilon+p}{\rho}\delta^{ij} \ .
\ee

\bibliographystyle{apsrev4-1}

\bibliography{biblio}

\end{document}